\documentclass[aps,pra,reprint,twocol-mn,floatfix,superscriptaddress,nofootinbib]{revtex4-1}

\usepackage[T1]{fontenc}
\usepackage{newtxtext}
\usepackage{amssymb}   
\usepackage[noamssymbols]{newtxmath}
\usepackage{graphicx,amsbsy,amsmath,bm,tikz}
\usepackage[version=4]{mhchem}
\usepackage{orcidlink}
\usepackage{tikzscale}
\usepackage{babel}
\usepackage{standalone}
\usepackage{xcolor}

\usepackage{dcolumn}   
\usepackage{mathtools,empheq}
\usepackage{hyperref}
\usepackage{booktabs}
\usepackage{float}
\usepackage{epstopdf}
\usepackage{siunitx} 
\usepackage{gensymb} 
\usepackage{braket}
\usepackage{bm} 
\usepackage{array} 
\newlength{\cfgcolw}
\newcolumntype{C}[1]{>{\centering\arraybackslash}m{#1}}
\newcommand{\et}{{\em et al.~}}
\newcommand{\integ}[3]{\int\displaylimits_{#1}^{#2}\! {{\rm d} }#3}

\newcommand{\qquestion}[1]{}
\usepackage[capitalise]{cleveref} 
\Crefname{figure}{Figure}{Figures}

\hypersetup{
    colorlinks=true,
    linkcolor=blue,
    filecolor=blue,
    urlcolor=blue,
    citecolor=blue,
}
\newcolumntype{C}[1]{>{\centering\arraybackslash}m{#1}}
\usepackage{tikz}
\usepackage{pgfplots}
\begin{document}

\title{
    Electron loss and target excitation in keV-energy proton collisions with 
    \ce{B} and \ce{C^+} 
}

\author{N.~W.~Antonio\,\orcidlink{0000-0003-3900-6197}}
\email{nick.antonio@curtin.edu.au}
\affiliation{Department of Physics and Astronomy, 
    Curtin University, 
    GPO Box U1987, 
    Perth, 
    WA 6845, 
    Australia
}

\author{I.~B. Abdurakhmanov\,\orcidlink{0000-0002-9844-4966}}
\affiliation{
    Pawsey Supercomputing Research Centre, 
    1 Bryce Avenue, 
    Kensington, 
    WA 6151, 
    Australia
}

\author{A. Espinosa-Gayosso\,\orcidlink{0000-0002-8869-3985}}
\affiliation{
    Pawsey Supercomputing Research Centre, 
    1 Bryce Avenue, 
    Kensington, 
    WA 6151, 
    Australia
}

\author{A.~S.~Kadyrov\,\orcidlink{0000-0002-5804-8811}}
\affiliation{Department of Physics and Astronomy, 
    Curtin University, 
    GPO Box U1987, 
    Perth, 
    WA 6845, 
    Australia
}

\date{\today}

\begin{abstract}
    The one-centre Coulomb-Sturmian convergent close-coupling method is applied 
    to proton collisions with the boron atom and singly charged carbon ion. 
    Here we report an update to our target-structure implementation, in which 
    configuration state functions are constructed using the method of 
    coefficients of fractional parentage. To assess the quality of the 
    structure models for the two targets, we present the excitation energies, 
    oscillator strengths, and dipole polarisabilities obtained from the present 
    configuration 
    interaction calculations. Cross sections for total and state-selective target 
    excitation and electron loss are calculated from 10 keV to 1 MeV. For both 
    systems, the total excitation cross section is found to be dominated by 
    excitation of the $2s$ subshell. This emphasises the importance of a 
    multi-electron description of the target in such scattering calculations. 
    Comparisons with previous theoretical and experimental data are presented and 
    discussed. In particular, we find that the present calculation for the 
    electron-loss cross section in \ce{$p$ + C^+} collisions is in good agreement 
    with the available measurements across the entire overlapping incident-energy 
    range.
\end{abstract}
	
\maketitle

\section{Introduction}
Obtaining accurate cross-section data for a wide range of ion–atom and 
ion–molecule collisions is an important endeavour. Such data underpin
many practical applications, including hadron therapy for cancer treatment 
\cite{Belkic_2010_47}, astrophysics \cite{Maurin_2026_1161}, and fusion plasma 
modelling \cite{Hill_2023_63}. Since the decision to change the material of the 
plasma-facing wall components of the ITER fusion reactor from \ce{Be} to \ce{W}
\cite{Pitts_2025_42}, there has been renewed interest in obtaining accurate 
cross-section data for various collisions involving \ce{B} and 
\ce{B}-containing molecules \cite{Kawate_2023_32,Mukherjee_2025_32}. This is 
because boron is introduced into the reactor to coat the plasma-facing 
components made of \ce{W} and thereby mitigate the formation of highly charged \ce{W} 
impurities arising from erosion of the reactor walls \cite{Bortolon_2019_19}. 
It is
introduced into the reactor through two main mechanisms, (i) via boronisation 
where the plasma-facing components are coated with a thin layer of boron 
through plasma discharges outside of operation, and (ii) through boron seeding 
during the operation of the reactor \cite{Snipes_2024_41}. As a result, 
\ce{B} and \ce{B}-containing species are expected to be present in the plasma, 
and therefore various ion collisions with these targets will occur 
\cite{Kawate_2022_62}. The emissions by species formed in their excited states 
during these collisions are used
to perform various diagnostics on the plasma. Such spectroscopic diagnostic
techniques include charge exchange recombination spectroscopy (CXRS) 
\cite{Isler_1994_36,McDermott_2021_61} and beam emission spectroscopy (BES)
\cite{Anderson_2000_42}. These diagnostic methods rely on having an accurate 
database of cross sections for all of the possible collisions that can occur in 
the plasma, including those involving \ce{B}.

Despite the importance of having accurate cross-section data for ion collisions 
involving \ce{B}, the data available in the literature for such collisions are 
scarce. On the theoretical front, Peach
\cite{Peach_1968_01,Peach_1970_03} applied the first-Born 
approximation (FBA) to calculate the cross section for ionisation in \ce{$p$ + B} 
collisions between 6 and 1522 keV. However, later it was discovered 
\cite{Peach_1971_04} that the results published in Refs. 
\cite{Peach_1968_01,Peach_1970_03} were in error by approximately a factor of 
two. To the best of our knowledge, the corrected data is no longer accessible.

Further work using perturbative methods was also performed to calculate 
the cross section for K-shell ionisation in high-energy \ce{$p$ + B} 
collisions. This includes the application of the binary-encounter 
approximation (BEA) by \citet{Garcia_1970_01}, which predicts a universal curve 
for the K-shell ionisation cross section that can be scaled as 
$\varepsilon_{1s}/ Z^2$, where $\varepsilon_{1s}$ is the K-shell 
ionisation energy and $Z$ is the charge of the target nucleus. 
The corrected plane-wave Born approximation (CPWBA) has also been 
applied to calculate the cross section for K-shell ionisation in \ce{$p$ + B} 
collisions by \cite{Basbas_1973_07,Ariyasinghe_1999_59}. Measurements of the 
K-shell ionisation cross section for 
\ce{$p$ + B} collisions have been reported by \citet{Ariyasinghe_1999_59},
\citet{Toburen_1972_05}, and \citet{Kobayashi_1976_40} over the incident-energy
range 400 keV to 2 MeV. Generally, the experimental data agree with one another 
across the overlapping incident energy ranges.
The Coulomb-deflection perturbed stationary state with relativistic correction
(CPSSR) method appears to be in better agreement with these 
measurements amongst the various predictions. Overall, K-shell ionisation is 
fairly well understood and will not be the focus of the present work. 

In terms of the target excitation and electron loss processes, there 
appears to be no theoretical or experimental data available in the literature 
for \ce{$p$ + B} collisions. The main goal of this work is to fill this 
gap. One of the likely reasons for the scarcity of data for 
ion collisions involving \ce{B} is the difficulty of treating the 
multi-electron nature of the target. At the very least, \ce{B} needs to be 
modelled as a quasi-three-electron system above the \ce{$1s^2$} core, which 
immediately complicates the problem \cite{Wang_2016_93}. Additionally, there is
strong coupling between the singly 
excited \ce{$2s^2 2p n\ell$} states to states with a vacancy (hole) in the 
inner-valence $2s$ subshell, such as those with configurations 
\ce{$2s 2p^2 n\ell$} and \ce{$2p^3 n\ell$}. When considering the K-shell 
ionisation process, a multi-electron treatment becomes even more challenging as 
it requires the inclusion of states with holes in the \ce{$1s$} shell.

Recently, we have extended the semi-classical one-centre implementation of the 
Coulomb–Sturmian convergent close-coupling (CS-CCC) method 
\cite{Antonio_2025_111} to treat ion 
collisions with arbitrary multi-electron targets \cite{Antonio_2025_112}. This 
was achieved by 
developing a new multi-electron structure package based on a spin-orbital 
representation, primarily using Coulomb–Sturmian functions to diagonalise the 
$N$-electron target Hamiltonian. To demonstrate its capability, we initially 
applied the method to antiproton collisions with the C atom. The use of an 
antiproton projectile significantly simplifies the collision dynamics, as at 
keV incident energies electron capture is negligible, thereby reducing the 
number of states required in the close-coupling expansion. However, extending 
such one-centre calculations to positively charged projectiles has historically 
been prohibitively expensive, due to the substantially larger number of states 
required in the close-coupling expansion.

In collaboration with the Pawsey Supercomputing Research Centre, we 
subsequently ported the multi-electron CS-CCC code to GPUs using the 
Heterogeneous-compute Interface for Portability (HIP) programming model and 
corresponding BLAS libraries \cite{hipblas}. This optimisation resulted in a 
performance 
improvement of approximately a factor of 600. These gains have enabled 
calculations involving $\sim 10^4$ states in the close-coupling expansion, 
making previously intractable problems computationally feasible. While the 
technical details of the code optimisation will be reported elsewhere 
\cite{Abdurakhmanov2026}, this 
development now allows us to consider collisions of positively charged 
projectiles with multi-electron targets, which is the focus of the present 
work.

In this work, we calculate the total and state-selective cross sections for 
target excitation and electron loss in \ce{$p$ + B} collisions in the 
incident-energy range 10 keV to 1 MeV. Here, we also perform calculations for 
\ce{$p$ + C^{+}} collisions. This is because \ce{C^{+}} is a \ce{B}-like target 
and previous FBA calculations and 
experimental measurements of the electron-loss cross section are available for
\ce{$p$ + C^+} collisions, which we can compare our results with. 
Considering this collision system not only allows us to compare our method with 
experimental data for the first time, but also further supports the quality of 
our results for \ce{$p$ + B} collisions, for which there is no experimental 
data available in the literature.

This paper is organised as follows. In Sec. \ref{sec:theory}, we briefly 
outline the multi-electron CS-CCC method used in this work. Details of the 
calculations, including choices of numerical parameters and basis size used in 
the final calculations, are given in Sec. \ref{sec:details}. The results of our
calculations are presented and discussed in Sec. \ref{sec:results}. Finally, we
summarise our findings and give an outlook in Sec. \ref{sec:conclusions}.

Atomic units are used throughout this work unless otherwise stated.

\section{Formalism}\label{sec:theory}
Details of the multi-electron CS-CCC approach have been 
described in previous work \cite{Antonio_2025_112}. As such, here we
primarily focus on the further improvements we have made to our treatment of the target
structure since the publication of our previous work. However, a brief 
outline of the close-coupling formalism is also given for completeness.

\subsection{Atomic structure}
We consider an $N$-electron atomic target and
construct a set of pseudostates using the configuration interaction (CI) 
method to collectively represent its bound and continuous spectrum. 
Here, we use the $LS$-coupling scheme, and therefore each pseudostate is 
characterised 
by its total orbital angular momentum quantum number $L$, the total spin 
quantum number $S$, and a set of additional quantum numbers 
required to uniquely specify the state which we denote $\Gamma$. Each 
pseudostate is expressed as a 
linear combination of configuration state functions (CSFs) as
\begin{align}
    \ket{\psi(\Gamma, L, S)} 
    &=
    \sum_{j} c^{(\Gamma, L, S)}_j \ket{\phi(\gamma_{j}, L, S)},
\end{align}
where $c^{(\Gamma, L, S)}_j$ is the CI coefficient corresponding to the $j$-th 
CSF $\ket{\phi(\gamma_{j}, L, S)}$. The label $\gamma_{j}$ denotes the set of 
quantum numbers required to uniquely specify the $j$-th CSF, in addition to $L$ 
and $S$.

The CSFs are eigenfunctions of the squared total orbital angular momentum 
operator $\bm{L}^2$, and the squared total spin operator $\bm{S}^2$ and 
corresponding projection operators $L_z$ and $S_z$ so that each 
pseudostate has well-defined magnitudes and $z$-components of the total orbital 
and spin angular momenta.
In previous work \cite{Antonio_2025_112}, we constructed CSFs using a linear 
combination of Slater determinants each with a corresponding angular 
coefficient. These coefficients were simply expressed as a product of 
Clebsch-Gordan coefficients obtained by sequentially coupling the orbital and 
spin angular momenta of the electrons in the order they were listed in the 
configuration. However, while this is a valid 
and fairly simple method to construct CSFs, it can very often lead to sets of 
linearly dependent or not properly antisymmetrised CSFs, which can cause 
issues in subsequent CI calculations. While for small atoms this can be handled 
in a completely numerical way by manually identifying and removing problematic 
CSFs from the basis, this is not ideal as it can often be difficult to identify 
those that need to be removed. The procedure becomes even more difficult when 
constructing a basis of CSFs for atoms with increasingly larger numbers of 
electrons. To overcome this issue, in this work we have updated our method and 
corresponding code for constructing CSFs following the well-established 
coefficients of fractional parentage method 
\cite{Cowan_1981_00,Fischer_2016_49}. For completeness, we briefly outline the 
method here. We start with some arbitrary electron configuration 
$(n_1 \ell_1)^{w_1} \ldots (n_m \ell_m)^{w_m}$, where $n_a$, $\ell_a$ and 
$w_a$ are the principal quantum number, orbital angular momentum quantum number 
and occupation number of the $a$-th subshell, respectively. 
Note that $\sum_{a=1}^m w_a = N$. We start by
constructing a CSF for a single subshell. This can be done recursively. To 
form a CSF for $w$ (for brevity we omit here the subscript referring to a 
particular subshell) equivalent electrons in a subshell, we express it as a 
linear combination of $(w-1)$-electron CSFs coupled to a single electron as 
follows
\begin{align}
    \ket{\phi((n \ell)^{w}, \gamma L S)}
    &=
    \sum_{\gamma' L' S'}
    (\ell^{w-1} \gamma' L' S' |\} \ell^{w} \gamma L S)
    \\ \nonumber
    &\times
    \ket{\phi([(n \ell)^{w-1}, \gamma' L' S'], (n \ell), \gamma L S)},
    \label{eq:csf_construction_subshell}
\end{align}
where $(\ell^{w-1} \gamma' L' S' |\} \ell^{w} \gamma L S)$ is the coefficient 
of fractional parentage, and quantum numbers $\gamma'$, $L'$ and $S'$ are the 
ones corresponding to the $(w-1)$-electron CSF. To calculate the coefficients of
fractional parentage, we have rewritten parts of the \texttt{ANG} library
by \citet{Fischer_1991_64} into a modern \texttt{Fortran} package that is 
integrated into our code. Ket
$\ket{\phi([(n\ell)^{w-1} \gamma' L' S'], (n \ell), \gamma L S)}$ is 
the state obtained by coupling a single equivalent electron to a CSF for $w-1$ 
equivalent electrons according to
\begin{align}
    &\ket{\phi([(n\ell)^{w-1} \gamma' L' S'], (n\ell), \gamma L S)}
    =
    \sum_{M' \Sigma' m \sigma}
    C_{L', M', \ell, m}^{L, M}
    \nonumber \\
    &\quad \times
    C_{S', \Sigma', 1/2, \sigma}^{S, \Sigma}
    \ket{\phi((n\ell)^{w-1}, \gamma' L' S')} 
    \wedge
    \ket{\chi_{n \ell m \sigma}},
\end{align}
where $C^{L, M}_{L', M', \ell, m}$ and $C^{S, \Sigma}_{S', \Sigma', 1/2, \sigma}$ 
are Clebsch-Gordan coefficients, $\ket{\chi_{n \ell m \sigma}}$ is a spin 
orbital of the single electron being coupled to the parent $(w-1)$-electron CSF, and 
$\wedge$ denotes the wedge product, which is used to ensure that the resulting 
state is properly antisymmetrised with respect to the exchange of any two 
electrons. Quantum numbers $m$ and $\sigma$ are the orbital and spin projection 
numbers of the single electron being coupled, respectively, and $M'$ and 
$\Sigma'$ are the total orbital and spin projection quantum numbers of the 
$(w-1)$-electron CSF, respectively. That leaves $M$ and $\Sigma$ as the total 
orbital and spin projection quantum numbers of the resulting $w$-electron CSF, 
respectively. 

We continue to apply the recurrence relation in Eq. 
\eqref{eq:csf_construction_subshell} until we reach the trivial $w = 1$ case. 
In fact, the $w = 2$ level is also fairly simple as the coefficients of 
fractional parentage for two equivalent electrons are equal to 1 when the sum of 
the total orbital and spin angular momenta of the two electrons is even, and 0 
otherwise \cite{Sobelman_1979_00}. Having constructed valid 
CSFs for each set of equivalent electrons that comprise the global $N$-electron 
configuration, we can then iteratively apply the addition of angular momenta to
couple the CSFs together to obtain the CSF corresponding to the global
electron configuration with quantum numbers $\gamma_j$, $L$ and $S$ as follows
\begin{align}
    \ket{\phi(\gamma_j, L, S)}
    &=
    \sum_{\mu_{j} \in \Lambda_{j}}
    C_{L_{1},M_{1},L_{2},M_{2}}^{L_{12},M_{12}}
    \ldots
    C_{L_{12\ldots m-1},M_{12\ldots m-1},L_{m},M_{m}}^{L,M}
    \nonumber \\ 
    &\times
    C_{S_{1},\Sigma_{1},S_{2},\Sigma_{2}}^{S_{12},\Sigma_{12}}
    \ldots
    C_{S_{12\ldots m-1},\Sigma_{12\ldots m-1},S_{m},\Sigma_{m}}^{S,\Sigma}
    \nonumber \\
    &\times
    \ket{\phi((n_1 \ell_1)^{w_1}, \gamma_1 L_1 S_1)} 
    \nonumber \\
    &\wedge \ldots \wedge
    \ket{\phi((n_m \ell_m)^{w_m}, \gamma_m L_m S_m)},
\end{align}
where 
\begin{align}
    \mu_{j} \equiv \{M_1, M_2, M_{12}, M_{3}, M_{123}, \ldots,
    \Sigma_1, \Sigma_2, \Sigma_{12}, \Sigma_{3}, \Sigma_{123},\ldots\}
\end{align}
is a complete set of projection quantum numbers of each subshell CSF as well 
as the intermediate projection quantum numbers of the coupled states, and 
$\Lambda_{j}$ is the set of all possible $\mu_{j}$.

To obtain the CI coefficients $c^{(\Gamma, L, S)}_j$ and corresponding energy 
eigenvalues $\varepsilon_{\Gamma, L, S}$, we diagonalise the $N$-electron 
target Hamiltonian $H_{\rm T}$ in the basis of CSFs, which is equivalent to 
solving the following generalised eigenvalue problem
\begin{align}
    \bm{H} 
    \bm{C}
    = 
    \bm{S} 
    \bm{C}
    \bm{\varepsilon},
    \label{eq:generalised_eigenvalue_problem}
\end{align}
where $\bm{H}$ is the Hamiltonian matrix with elements $H_{ij} = 
\bra{\phi(\gamma_i, L, S)} H_{\rm T} \ket{\phi(\gamma_j, L, S)}$, $\bm{S}$ is the 
overlap matrix with elements $S_{ij} = \braket{\phi(\gamma_i, L, S) | 
\phi(\gamma_j, L, S)}$, $\bm{C}$ is the matrix of CI coefficients and 
$\bm{\varepsilon}$ is the diagonal matrix of energy eigenvalues. Note that the 
CSFs are orthogonal with respect to parity, $L$ and $S$ and therefore we solve
Eq. \eqref{eq:generalised_eigenvalue_problem} separately for each symmetry 
block defined by these quantum numbers.

\subsection{Semi-classical convergent close-coupling method}
The coupled-channel equations are derived in the position representation.
To this end, we define $\bm{X}$ as the set of all spatial and spin coordinates 
of the $N$ electrons which comprise the target. This way we write the 
pseudostate wavefunctions as follows
\begin{align}
    \braket{\bm{X} | \psi(\Gamma, L, S)} 
    &=
    \psi_{\Gamma L S}(\bm{X})
        \equiv
    \psi_{\alpha}(\bm{X}).
    \label{}
\end{align}
Here and what follows we collect all quantum numbers into a single index 
$\alpha$. We also define $\bm{\sigma}$ as 
the position of the incoming projectile ion relative to the centre of mass of 
the target, and $\bm{R}$ as the position of the projectile ion relative to the 
target nucleus. We express our trial scattering wavefunction, 
$\Psi^{(+)}_{i}(\bm{\sigma}, \bm{X})$, as follows
\begin{align}
        \Psi^{(+)}_{i}(\bm{\sigma}, \bm{X})
        =
        \sum_{\alpha = 1}^{N_{\rm T}}
        F_{\alpha}(\bm{\sigma})
        {\rm e}^{i\bm{k}_{\alpha} \cdot \bm{\sigma}}
        \psi_{\alpha}(\bm{X}), 
        \label{eq:trial_wavefunction}
\end{align}
where superscript $(+)$ denotes that the trial wavefunction satisfies the 
appropriate outgoing wave boundary conditions, $i$ is the index of the initial 
state of the target, and $N_{\rm T}$ is the total number of pseudostates 
included in the expansion. Vector $\bm{k}_{\alpha}$ is the momentum of 
the projectile ion relative to the target in the channel $\alpha$. Coefficients 
$F_{\alpha}(\bm{\sigma})$ are initially unknown and need to be determined in 
order to calculate the scattering amplitudes and cross sections. To do this we 
use the Petrov-Galerkin method \cite{Reddy_1993_00} to solve the 
time-independent Schr\"odinger equation of the collision system. This yields
the following set of equations
\begin{align}
    \integ{}{}{\bm{X}}
    \psi^{*}_{\alpha'}(\bm{X})
    {\rm e}^{-i\bm{k}_{\alpha'} \cdot \bm{\sigma}}
    \left(H - E\right)
    \Psi^{(+)}_{i}(\bm{\sigma}, \bm{X})
     = 
     0,
     \label{eq:petrov_galerkin}
\end{align}
where $H$ is the total Hamiltonian and $E$ is the total energy of 
the scattering system and $\alpha'$ ranges from 1 to $N_{\rm T}$. Simplifying Eq. 
\eqref{eq:petrov_galerkin} with no approximation leads to a set of fully 
coupled differential equations which is numerically intractable to solve. 
As we are working in the collision energy regime of 1 keV and above, the 
semi-classical approximation is valid. This assumes 
that we can treat the motion of the incoming projectile ion classically, while 
leaving the electron dynamics described quantum mechanically. If we centre the 
target nucleus at the origin in the laboratory frame and take the projectile 
ion to be incident along the $z$-axis, then the projectile ion 
trajectory is given by $\bm{R} = \bm{b} + z\bm{\hat{z}}$, where $\bm{b}$ is the 
impact parameter. Using the semi-classical approximation and also setting
$F_{\alpha}(\bm{\sigma}) \approx F_{\alpha}(\bm{R})$ leads to the 
following system of coupled-channel (CC) equations
\begin{align}
    iv \partial_{z} F_{\alpha'}(z,\bm{b})
    &=
    \sum_{\alpha = 1}^{N_{\rm T}}
    F_{\alpha}(z,\bm{b})
    \braket{\psi_{\alpha'}| \overline{V} | \psi_{\alpha} }
    {\rm e}^{i(k_{\alpha} - k_{\alpha'}) z},
    \label{eq:coupled_channel_equations}
\end{align}
which can be readily solved numerically subject to some initial boundary 
condition. Here, $\overline{V}$ is the 
interaction potential between the projectile ion and the target, and $v$ is the 
velocity of the projectile ion. Symbol $\partial_{z}$ denotes the partial
derivative with respect to $z$. For details on how we evaluate the scattering 
matrix elements we refer the reader to our previous work 
\cite{Antonio_2025_112}. Eq. \eqref{eq:coupled_channel_equations} is solved for 
each impact parameter with the initial condition
\begin{align}
    F_{\alpha}(z \to -\infty, \bm{b}) = \delta_{\alpha i},
    \label{}
\end{align}
which implies that before the collision the target atom is in state $i$.

All cross sections we present in this work are averaged over the initial
magnetic sub-states of the target, and summed over the final magnetic 
sub-states.

\section{Details of the calculations}\label{sec:details}

\subsection{Target structure model}
One of the most difficult aspects of modelling collisions with multi-electron 
targets like \ce{B} and \ce{C^+} in terms of target structure is constructing a 
pseudostate basis which not only accurately represents a few to several low-lying 
bound states of the target, but also provides a good representation of the 
target's continuous spectrum. This must be achieved while keeping the size of 
the pseudostate basis manageable for subsequent scattering calculations.
This is in contrast to the challenges faced in the field of quantum chemistry where 
the goal is to obtain highly accurate results for only a few of the low-lying 
bound states of the target
\cite{Bubin_2011_83,Bubin_2017_118}.

In order to obtain a structure model suitable for our needs, we need to include 
a sufficient number of electron configurations 
to capture most of the relevant electron correlation effects, while also 
choosing a set of radial orbital basis functions that are sufficiently flexible 
to represent the entire spectrum of the target. Beginning with the latter, we start with a basis of Coulomb-Sturmian (CS) functions 
\cite{Rotenberg_1962_19}. We then perform a Hartree-Fock (HF) calculation using 
the code developed by \citet{Fischer_2019_00} to obtain $1s$, $2s$ and $2p$ orbitals 
optimised for the ground state of the target. The corresponding CS functions 
with the same quantum numbers are then replaced with these optimised HF 
orbitals. Additionally, in the case of \ce{B}, we also replace the CS $3s$ 
orbital with a HF $3s$ orbital optimised for the \ce{$2s^2 3s$ ^2S} state of 
the target. The falloff parameter of the remaining CS functions, $\zeta$, is 
set to 1.0 in all cases. 

\begin{table}[t]
    \caption{
        Electron configurations included in the \ce{B} and \ce{C^+} 
        target structure models.
    }
    \label{tab:B_C+_configs}
    \begin{ruledtabular}
    \begin{tabular}{ll}
        \textbf{Configuration type} & \textbf{Details / Range} \\
        \hline
        $2s^2 nl$     & $n = 2\text{--}20 \quad l = 0\text{--}8$ \\
        $2s 2p nl$    & $n = 3\text{--}20 \quad l = 0\text{--}6$ \\
        $2p^2 nl$     & $n = 2\text{--}20 \quad l = 0\text{--}4$ \\
        $2s 3p nl$    & $n = 3\text{--}12 \quad l = 0\text{--}2$ \\
        $2s 3l nl$    & $n = 3\text{--}8 \quad l = 0, 2$ \\
        $2s 4l nl$    & $n = 4\text{--}8 \quad l = 0, 2$ \\
        $2p nl^2$     & $n = 3\text{--}5 \quad l = 0\text{--}3$ \\
        $2p 3p np$    & $n = 3\text{--}14$ \\
        $2p np n'd$   & $n = 2\text{--}5 \quad n' = 3\text{--}18$ \\
        $ns n'p n''p$ & $n, n' = 2\text{--}4 \quad n'' = 2\text{--}8$ \\
        $2p^3$, $2s 2p 3p$, $2s 3p^2$ & Discrete configurations \\
    \end{tabular}
    \end{ruledtabular}
\end{table}

To not only obtain an accurate target structure model but to also achieve 
convergence in the subsequent scattering calculations, we need to include a 
large set of different electron configurations to construct a set of CSFs.
These configurations are listed in Table \ref{tab:B_C+_configs}. The excitation 
energies of some of the low-lying states of \ce{B} and \ce{C^+} obtained from our 
CI calculations are presented in Tables \ref{tab:excitation_energies_B} and 
\ref{tab:excitation_energies_C+}, respectively, together with the corresponding 
values from the NIST Atomic Spectra Database \cite{NIST_ASD}. For \ce{B}, we 
also include the CI results of \citet{Wang_2016_93}, who employed a hybrid 
spin-orbital basis comprising Hartree-Fock, multiconfigurational Hartree-Fock, 
and B-spline orbitals. Overall, our results are in good agreement with the NIST 
data for both targets, with most excitation energies differing by about 1--2\% 
or less. In the case of \ce{B}, we also find good agreement with the results of 
\citet{Wang_2016_93} for most states. However, for the \ce{^2S} states, our 
results slightly deviate from the NIST data. The largest discrepancy occurs for 
the \ce{$2s^2 4s$ ^2S} state. The origin of this deviation is not entirely clear, 
but it is likely that the radial basis used in the present calculations is not 
optimal for describing this state. We note, however, that \citet{Wang_2016_93} 
employed term-dependent radial orbitals optimised for particular terms, which 
enables closer agreement with the NIST values for the states listed in Table 
\ref{tab:excitation_energies_B}. Their calculations therefore provide an 
important benchmark for assessing the quality of the present structure model.

\begin{table*}[t]
    \caption{
        Excitation energies of the B atom. The present CI calculations are 
        shown alongside the results of \citet{Wang_2016_93} and the NIST Atomic
        Spectra Database \cite{NIST_ASD}. The percentage difference between the
        present results and the NIST values is also given in the last column.
        All energies reported are in atomic units (a.u.).
    }
    \label{tab:excitation_energies_B}
    \begin{ruledtabular}
    \begin{tabular}{llcccc}
        \textbf{State} & \textbf{Term} & \textbf{Wang \et} \cite{Wang_2016_93} & \textbf{Present} & \textbf{NIST \cite{NIST_ASD}} & \textbf{Percentage difference} \\
        \hline
        $2s^2 2p$ & \ce{^2P^{\rm o}} & 0.0000 & 0.0000 & 0.0000 & 0.000  \\
        $2s^2 3s$ & \ce{^2S}         & 0.1807 & 0.1738 & 0.1824 & $-4.71$ \\
        $2s 2p^2$ & \ce{^2D}         & 0.2183 & 0.2151 & 0.2181 & $-1.38$ \\
        $2s^2 3p$ & \ce{^2P^{\rm o}} & 0.2195 & 0.2202 & 0.2215 & $-0.59$ \\
        $2s^2 3d$ & \ce{^2D}         & 0.2471 & 0.2457 & 0.2495 & $-1.52$ \\
        $2s^2 4s$ & \ce{^2S}         & 0.2481 & 0.2700 & 0.2506 & 7.74  \\
        $2s^2 4p$ & \ce{^2P^{\rm o}} & 0.2608 & 0.2626 & 0.2633 & $-0.27$ \\
        $2s^2 4d$ & \ce{^2D}         & 0.2710 & 0.2695 & 0.2734 & $-1.43$ \\
        \hline
        Ionisation limit & & 0.4118  & 0.3033 & 0.3049 & $-0.52$ \\
        \hline
        $2s 2p^2$ & \ce{^2P}        & --     & 0.3289 & 0.3304 & $-0.45$ \\
    \end{tabular}
    \end{ruledtabular}
\end{table*}

It is interesting to note the considerable differences in the ordering of the 
states between the \ce{B} and \ce{C^+} targets. For \ce{C^{+}}, we see that the 
first three excited states are inner-valence excited states with a hole in the 
$2s$ subshell, while for \ce{B} these types of states, except for the \ce{$2s 
2p^2$ ^2P} state, are above the ionisation limit. Also in the case of \ce{C^+}, we 
see the presence of bound $2p^3$ states, which are not present in the case of 
\ce{B}. These differences are a consequence of the strong attraction to the 
\ce{C^{6+}} nucleus, which leads to a stronger binding of the electrons.
Such differences in the target structure and resulting spectrum are expected to have a 
significant effect on the collision dynamics which is seen in Sec. 
\ref{sec:results}. This emphasises the importance of having a good description 
of the target structure to be able to obtain reliable results for the 
collisions with these targets. 

\begin{table*}[t]
    \caption{
        Excitation energies of the C$^+$ ion. The present CI calculations 
        are shown alongside the NIST values \cite{NIST_ASD}. The percentage 
        difference between the present results and the NIST values is also 
        given in the last column.
        All energies reported are in atomic units (a.u.).
    }
    \label{tab:excitation_energies_C+}
    \begin{ruledtabular}
    \begin{tabular}{llccc}
        \textbf{State} & \textbf{Term} & \textbf{Present} & \textbf{NIST \cite{NIST_ASD}} & \textbf{Percentage difference} \\
        \hline
        $2s^2 2p$ & \ce{^2P^{\rm o}} & 0.0000 & 0.0000 & $0.00$ \\
        $2s 2p^2$ & \ce{^2D}         & 0.3401 & 0.3414 & $-0.38$ \\
        $2s 2p^2$ & \ce{^2S}         & 0.4466 & 0.4397 & $+1.57$ \\
        $2s 2p^2$ & \ce{^2P}         & 0.5088 & 0.5040 & $+0.95$ \\
        $2s^2 3s$ & \ce{^2S}         & 0.5192 & 0.5310 & $-2.22$ \\
        $2s^2 3p$ & \ce{^2P^{\rm o}} & 0.5882 & 0.6002 & $-1.99$ \\
        $2s^2 3d$ & \ce{^2D}         & 0.6718 & 0.6632 & $+1.29$ \\
        $2p^3$    & \ce{^2D^{\rm o}} & 0.6964 & 0.6856 & $+1.57$ \\
        $2s^2 4s$ & \ce{^2S}         & 0.7078 & 0.7164 & $-1.20$ \\
        $2s^2 4p$ & \ce{^2P^{\rm o}} & 0.7502 & 0.7405 & $+1.31$ \\
        $2s^2 4d$ & \ce{^2D}         & 0.7749 & 0.7660 & $+1.16$ \\
        $2p^3$    & \ce{^2P^{\rm o}} & 0.7895 & 0.7688 & $+2.69$ \\
        $2s^2 4f$ & \ce{^2F^{\rm o}} & 0.7843 & 0.7699 & $+1.87$ \\
        \hline
        Ionisation limit & & 0.8953 & 0.8961  & $-0.09$ \\
    \end{tabular}
    \end{ruledtabular}
\end{table*}

As another check of the quality of our structure models we also present a set 
of oscillator strengths for some of the low-lying dipole-allowed transitions 
in Tables \ref{tab:oscillator_strengths-B} and \ref{tab:oscillator_strengths-C+}
for \ce{B} and \ce{C^+}, respectively. For \ce{B}, we also 
present the corresponding results of \citet{Wang_2016_93} for comparison, while 
for both targets we compare with the NIST values \cite{NIST_ASD}. Generally, 
the present oscillator strengths are in good agreement with the NIST values and 
in the case of \ce{B} also with the results of Wang \et, which gives us 
confidence in the quality of our models. 

\begin{table*}[htpb]
    \caption{
        Oscillator strengths of the \ce{B} atom. The present CI calculations 
        are shown alongside the results of \citet{Wang_2016_93} and the NIST 
        Atomic Spectra Database \cite{NIST_ASD}.
    }
    \label{tab:oscillator_strengths-B}
    \begin{ruledtabular}
    \begin{tabular}{lccc}
        \textbf{Transition} & \textbf{\citet{Wang_2016_93}} & \textbf{Present} & \textbf{NIST \cite{NIST_ASD}} \\ 
        \hline
        \ce{$2s^2 2p$ ^2 P^{\rm o}} $\to$ \ce{$2s^2 3s$  ^2 S} & 0.0803 & 0.0814 & 0.0785 \\ 
        \ce{$2s^2 2p$ ^2 P^{\rm o}} $\to$ \ce{$2s 2p^2$  ^2 D} & --     & 0.0385 & 0.0471 \\ 
        \ce{$2s^2 2p$ ^2 P^{\rm o}} $\to$ \ce{$2s^2 3d$  ^2 D} & 0.1720 & 0.1724 & 0.1700 \\
        \ce{$2s^2 2p$ ^2 P^{\rm o}} $\to$ \ce{$2s^2 4s$  ^2 S} & 0.0162 & 0.0182 & 0.0154 \\ 
        \ce{$2s^2 2p$ ^2 P^{\rm o}} $\to$ \ce{$2s^2 4d$  ^2 D} & 0.0762 & 0.0772 & 0.0723 \\ 
        \ce{$2s^2 2p$ ^2 P^{\rm o}} $\to$ \ce{$2s 2p^2$  ^2 P} & --     & 0.5604 & 0.5847 \\ 
    \end{tabular}
    \end{ruledtabular}
\end{table*}

\begin{table}[htpb]
    \caption{
        Oscillator strengths of the \ce{C^+} ion. The present CI calculations 
        are shown alongside the NIST values \cite{NIST_ASD}.
    }
    \label{tab:oscillator_strengths-C+}
    \begin{ruledtabular}
    \begin{tabular}{lcc}
        \textbf{Transition} & \textbf{Present} & \textbf{NIST \cite{NIST_ASD}} \\ 
        \hline
        \ce{$2s^2 2p$ ^2 P^{\rm o}} $\to$ \ce{$2s 2p^2$ ^2 D} & 0.1230 & 0.1281 \\ 
        \ce{$2s^2 2p$ ^2 P^{\rm o}} $\to$ \ce{$2s 2p^2$ ^2 S} & 0.1305 & 0.1183 \\ 
        \ce{$2s^2 2p$ ^2 P^{\rm o}} $\to$ \ce{$2s 2p^2$ ^2 P} & 0.5016 & 0.4981 \\ 
        \ce{$2s^2 2p$ ^2 P^{\rm o}} $\to$ \ce{$2s^2 3s$ ^2 S} & 0.0102 & 0.0163 \\ 
        \ce{$2s^2 2p$ ^2 P^{\rm o}} $\to$ \ce{$2s^2 3d$ ^2 D} & 0.3659 & 0.3330 \\ 
    \end{tabular}
    \end{ruledtabular}
\end{table}

The dipole polarisabilities of the ground states of \ce{B} and \ce{C^+} are 
also calculated as a final check of the quality of our structure models. For 
\ce{B} we obtain a value of 20.86 $a_0^3$ which differs by about 1\% from the 
accepted value of 20.5 $\pm$ 0.1 $a_0^3$ \cite{Schwerdtfeger_2019_117}. In the 
case of \ce{C^{+}} we obtain a dipole polarisability of 5.61 $a_0^3$. 
Compared with the coupled-cluster single-double (CCSD) calculation of 
\citet{Wang_2021_75}, which gives a value of 5.77 $a_0^3$, the present result 
differs by about 1\%.

\subsection{Convergence tests and numerical parameters}
As in all convergent close-coupling calculations, we ensure that the cross 
sections of interest converge to within a few percent with respect to 
increasing the size of the pseudostate basis used to expand the trial 
scattering wavefunction \eqref{eq:trial_wavefunction}. The way the CS-CCC 
program works is that we only control the size of the pseudostate basis through 
the electron configurations that we include in the CI calculations. This way 
each term symbol that a particular electron configuration can give rise to, 
will result in a pseudostate with that term symbol being included in the 
expansion of the scattering wavefunction. Therefore we simply establish 
convergence with the maximum single electron orbital quantum number 
$l_{\rm max}$ and the maximum principal quantum number $n_{\rm max}$ of the 
orbitals used in the construction of the CSFs, whose values are suggested by 
the ranges given in Table \ref{tab:B_C+_configs}. In this work, due to the 
significant computational 
performance improvements of the GPU implementation of the CS-CCC method, we are 
able to include the entire set of pseudostates generated from the CI 
calculations in the close-coupling calculations, which was previously
infeasible. For both collisions with \ce{B} and \ce{C^+}, we include a 
total of 16000 pseudostates in the close-coupling calculations in order to 
obtain converged results for the cross sections presented below. 
This is an order of magnitude larger than the number of pseudostates we were
able to include in previous CS-CCC calculations for multi-electron targets 
\cite{Antonio_2025_112}. Such a large basis size is necessary to achieve 
convergence for the cross sections of interest in this work, as here we are 
considering proton collisions with \ce{B} and \ce{C^{+}}. Being inherently a 
two-centre problem, we typically need a much larger pseudostate basis to 
achieve convergence using a single-centre formalism 
\cite{Abdurakhmanov_2020_53}.

The coupled-channel equations in Eq. \eqref{eq:coupled_channel_equations} are 
solved using the fourth-order Runge-Kutta method on a discretised $z$-grid 
within the interval $[-z_{\rm max}, z_{\rm max}]$. For both targets and all
incident energies at which calculations are performed, we set $z_{\rm max} = 125$ 
a.u. and discretise the $z$-grid into 2400 exponentially distributed points. 
The exponential distribution of the points allows us to have a higher density 
of points around $z=0$ where the most variation in the coefficients 
$F_{\alpha}(z, \bm{b})$ is expected to occur, while having a lower density of 
points at large $|z|$ where the coefficients vary more slowly. 
We have established that the results presented in this work are converged with 
respect to varying $z_{\rm max}$ and the number of points discretising the 
$z$-grid to three significant figures. To ensure the stability of our solution
to Eq. \eqref{eq:coupled_channel_equations} for each incident energy and impact 
parameter, we compute $|\Psi^{(+)}_{i}|^2$ at each step of the Runge-Kutta 
propagation to ensure it is equal to 1 within a tolerance of $10^{-3}$. 

The integration over the impact parameter is performed using a 
piecewise-uniform grid on $[0,b_{\max}]$ where $b_{\max}$ is the maximum impact
parameter value. The grid is divided into three
linearly spaced segments whose density decreases in each successive segment.
The value of $b_{\max}$ is chosen such that the impact-parameter weighted 
transition probabilities, for all processes of interest have fallen by at least 
three orders of magnitude from their maximum values by $b=b_{\max}$. For 
\ce{$p$ + B} collisions, we use $b_{\max}=40$ a.u. for incident energies from 
10 to 100 keV, $b_{\max}=80$ a.u. for energies from 100 keV up to 500 keV,
and $b_{\max}=100$ a.u. above 500 keV. For \ce{$p$ + C^+} collisions, we use
$b_{\max}=40$ a.u. from 10 to 300 keV and $b_{\max}=60$ a.u. above 300 keV.
For a particular transition to final state $f$ with probability $P_{f}(b)$, the 
corresponding cross section is
\begin{align}
    \sigma_{i\to f} = 2\pi \int_0^\infty bP_{f}(b)\,{\rm d}b,
\end{align}
where $P_{f}(b)$ is obtained from the coefficient $F_{f}$ 
in the limit $z \to +\infty$ according to
\begin{align}
    P_{f}(b) = |F_{f}(z \to +\infty, \bm{b}) - \delta_{fi}|^2.
    \label{}
\end{align}
To ensure the accuracy of the impact-parameter integration, in each calculation
we repeat the integration for all fully state resolved transitions after 
removing every second $b$-grid point and compare the results to make sure they 
are the same.

\begin{figure}
    \centering
    \includegraphics[width=\columnwidth]{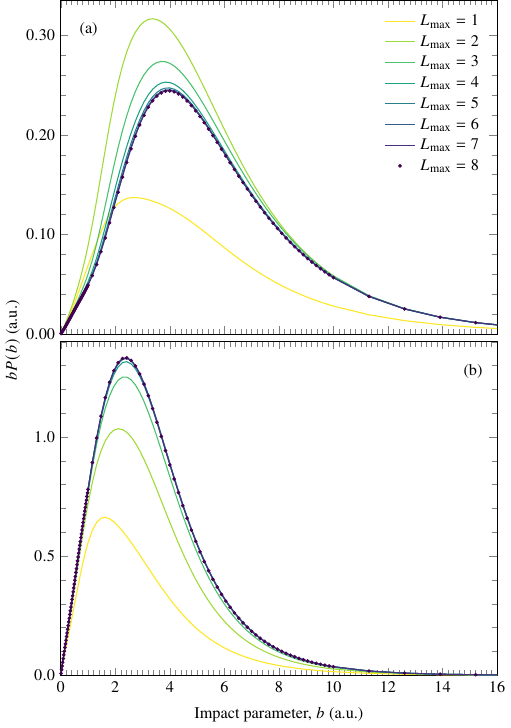}
    \caption{
        Convergence of the weighted transition probabilities 
        for (a) total excitations and (b) total electron loss
        in \ce{$p$ + B} collisions at 50 keV incident energy with 
        respect to $L_{\rm max}$, the largest total orbital angular momentum of  
        the target states included in the close-coupling expansion. Note that 
        only a lower part of the impact 
        parameter range is shown in each panel to highlight the most 
        relevant region of the transition probabilities. Legend in panel (a) 
        applies to both panels.
    }
\label{fig1}
\end{figure}

\begin{figure*}[t]
    \centering
    \hspace*{-1cm}
    \includegraphics{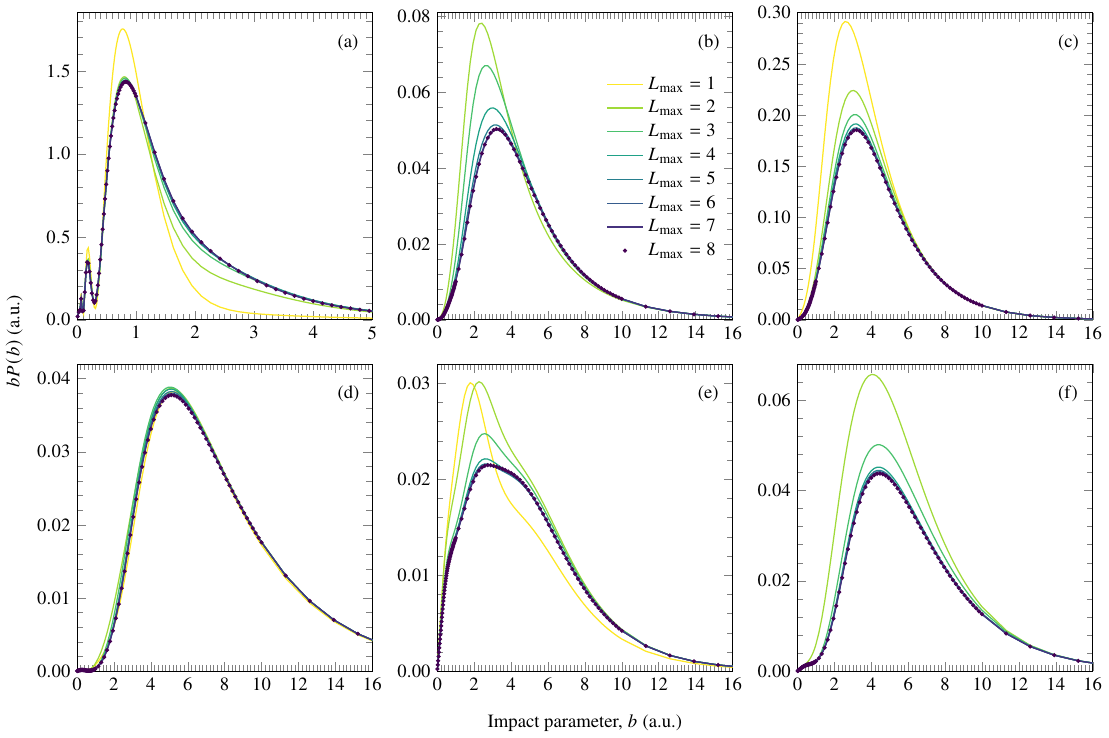}
    \caption{
        Convergence of the weighted transition probabilities 
        for (a) elastic scattering and excitations of the (b) \ce{$2s 2p^2$ ^2D}, 
        (c) \ce{$2s 2p^2$ ^2P}, (d) \ce{$2s^2 3s$ ^2S}, (e) \ce{$2s^2 3p$ 
        ^2P^{\rm o}}, and (f) \ce{$2s^2 3d$ ^2D} states of 
        \ce{B} in \ce{$p$ + B} collisions at 50 keV incident energy with 
        respect to $L_{\rm max}$, the largest total orbital angular momentum of  
        the target states included in the close-coupling expansion. Note that 
        only a lower part of the impact 
        parameter range is shown in each panel to highlight the most 
        relevant region of the transition probabilities. Legend in panel (b) 
        applies to all panels.
    }
\label{fig2}
\end{figure*}

Figures~\ref{fig1} and \ref{fig2} provide representative examples of 
both the choice of impact-parameter grid and the convergence of the weighted 
transition probabilities with respect to 
$L_{\max}$, the largest total orbital angular momentum of the target 
states included in the close-coupling expansion of the trial scattering 
wavefunction. Figure \ref{fig1} shows $bP(b)$ for total excitation and 
electron loss as a function of $b$ for \ce{$p$ + B} collisions at an incident 
energy of 50 keV. Similarly, Fig. \ref{fig2} shows $bP(b)$ for elastic 
scattering and excitations of the \ce{$2s 2p^2$ ^2D}, \ce{$2s 2p^2$ ^2P}, 
\ce{$2s^2 3s$ ^2S}, \ce{$2s^2 3p$ ^2P^{\rm o}}, and \ce{$2s^2 3d$ ^2D} states 
of \ce{B} in \ce{$p$ + B} collisions also at 50 keV incident energy.
Because the ground state of \ce{B} has $L=1$, the convergence sequence begins 
at $L_{\max}=1$. At each value of $L_{\max}$, all target states with 
$L\leq L_{\max}$ and both parities are included. The dots representing the 
$L_{\max}=8$ results mark the actual impact parameters used in the 
cross-section calculations presented in the next section.
The transition probabilities vary smoothly between neighbouring
grid points, and the oscillatory behaviour of the elastic-scattering
probability is well resolved by the denser low-$b$ segment. 
We see that the transition probabilities for all processes of interest have
converged very well with respect to $L_{\rm max}$. This is indicated by the fact 
that from about $L_{\rm max} > 4$ the curves for all processes of interest are 
essentially on top of each other. Convergence with respect to 
$L_{\rm max}$ is displayed here, because establishing convergence with respect 
to this basis parameter is typically the slowest and most challenging to 
achieve. Note that while we
only present convergence results for \ce{$p$ + B} collisions at 50 keV, we have
performed similar checks at a number of representative incident energies for both \ce{$p$ + 
B} and \ce{$p$ + C^+} collisions to ensure the reliability of the results 
presented in the following section.

\section{Results and discussion} \label{sec:results}
Throughout this section we present our results as points connected by straight 
lines to guide the eye of the reader. The points indicate the incident energies 
at which we performed our calculations.

\subsection{\ce{$p$ + B} collisions}
\begin{figure}[t]
    \centering
    \includegraphics[width=\columnwidth]{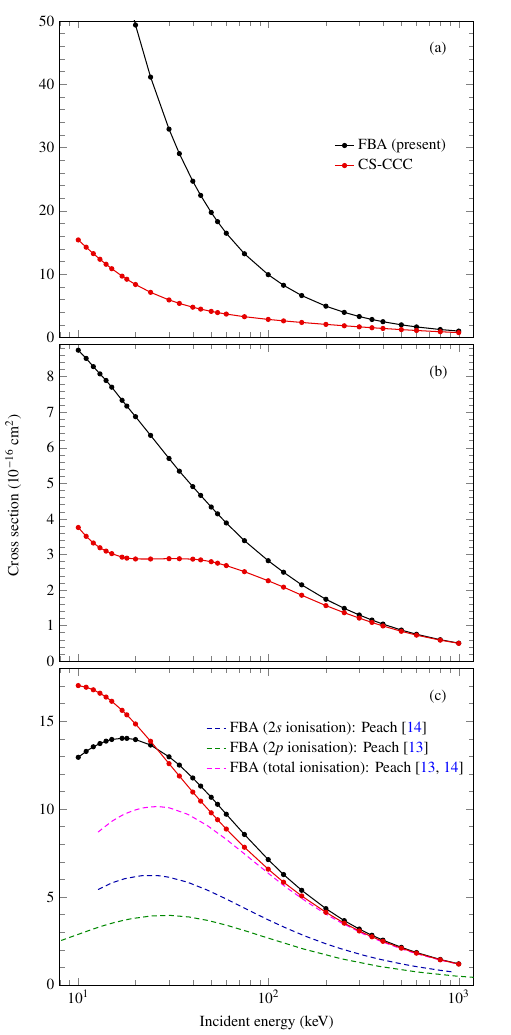}
    \caption{
        Cross sections for (a) elastic scattering, (b) total excitation 
        and (c) electron loss in \ce{$p$ + B} 
        collisions. Present calculations using the CS-CCC and FBA methods are 
        shown in all panels. In the lower panel the FBA cross sections for 
        ionisation of the $2s$ and $2p$ subshells by 
        \citet{Peach_1968_01,Peach_1970_03}, as well as the sum of these two 
        contributions are also shown.
    }
    \label{fig3}
\end{figure}

The CS-CCC and present FBA results for the cross sections of elastic 
scattering, total excitation and electron loss in \ce{$p$ + B} collisions are 
shown in panels (a), (b) and (c) of Figure \ref{fig3}, respectively. 
The CS-CCC results for all three processes appear to fall off monotonically 
with the collision energy from their values at the lowest energy of 10 keV 
considered in this work. As expected at sufficiently high incident energies, 
the FBA results in all three cases converge to the corresponding CS-CCC ones. 
Looking towards the lower incident energies we see varying degrees of deviation 
between the FBA and CS-CCC results for the different processes. The biggest 
difference between the two approaches is observed for the elastic-scattering 
cross section. To not suppress the CS-CCC results in panel (a), we do not show 
the corresponding FBA ones in full. However, we note that at 10 keV 
the cross section for elastic scattering obtained in the FBA is 
$9.87\times 10^{-15} {\rm cm}^2$, i.e. about 5 times larger than the 
corresponding CS-CCC result. The next largest 
difference between the two approaches is observed for the total excitation 
cross section where the FBA result is about 2 times larger than the CS-CCC 
result at 10 keV. The smallest difference between the two approaches is 
observed for the electron-loss cross section where the largest difference 
between the two methods across the entire incident energy range is about 30\%. 
Interestingly, at low energies the FBA results for electron loss are smaller 
than the CS-CCC ones. At incident energies between 20 and 40 keV, the CS-CCC 
total-excitation cross section shown in panel (b) of Fig. \ref{fig3}
exhibits a plateau that is absent in the FBA results. The origin of this 
feature becomes clear from the state-resolved excitation cross sections 
discussed below. At this stage, however, we note only that it must arise from 
coupling between the different channels in the close-coupling equations, which 
is why it is not present in the FBA results.
All of these differences, not only in magnitude but also in the functional form 
of the cross sections at lower incident energies, emphasise the importance of 
accounting for the coupling between channels to properly describe the collision 
dynamics in this energy regime. 

In panel (c) of Fig. \ref{fig3} we compare our calculated electron-loss 
cross section with the FBA cross sections for ionisation of the $2s$ and $2p$ 
subshells by \citet{Peach_1968_01,Peach_1970_03}, as well as the sum of these 
two contributions. Note that this summed total ionisation cross section is not 
explicitly presented in the original papers of Peach, but we have interpolated 
their results for the $2s$ and $2p$ ionisation cross sections onto a common 
incident energy grid and then summed the two contributions to enable comparison 
with our total electron-loss cross section. Additionally, we have accounted for 
the approximate factor of 2 error noted by Peach in Ref. \cite{Peach_1971_04}. 
To this end, all cross sections from Peach presented in Fig. \ref{fig3} have been 
divided by 2.
The FBA results in Refs. \cite{Peach_1968_01,Peach_1970_03} were 
calculated using the independent-particle approximation, in which the initial 
and residual ionic states are represented by a single CSF with optimised HF 
orbitals, while the ejected electron is described by an 
undistorted Coulomb continuum wave. Within this approximation, the $2p$- and 
$2s$-ionisation channels can be treated separately by choosing the active bound 
electron to be either the $2p$ or $2s$ electron and coupling to the 
appropriate residual-ion configuration, namely the \ce{$1s^2 2s^2$ ^1S} state 
of \ce{B^+} for $2p$ ionisation and the 
\ce{$1s^2 2s 2p$ ^3P^{\rm o}}/\ce{$1s^2 2s 2p$ ^1P^{\rm o}} states of \ce{B^+} 
for $2s$ ionisation. Because the transition amplitudes in the
FBA calculations do not contain channel-coupling contributions, they do not 
include any contributions from electron capture. Therefore, although we refer 
to the present FBA results as electron-loss cross section for consistency with 
the CS-CCC calculations, they actually represent the ionisation cross 
section. Above 300 keV, there is excellent agreement between the 
total FBA ionisation cross section of Peach and our calculated electron-loss 
cross section obtained using both the CS-CCC and FBA approaches. This is 
expected because, at incident energies of 300 keV and above, the contribution 
of electron capture to the total electron-loss cross section is negligible, so 
the total electron-loss cross section is essentially equal to the total 
ionisation cross section. At lower incident energies, however, the total 
ionisation cross section obtained by Peach in the FBA is significantly smaller 
than our corresponding FBA results. 
This difference could be due to the approximations used in their target 
structure as well as the treatment of the interaction between the ejected 
electron and residual target ion as purely Coulombic. In our FBA, on the other 
hand, the target structure is very accurate and the interaction of the ejected 
electron and the residual ion takes into account the multielectron nature of 
the ion. 

\begin{figure}[t]
    \centering
    \includegraphics[width=\columnwidth]{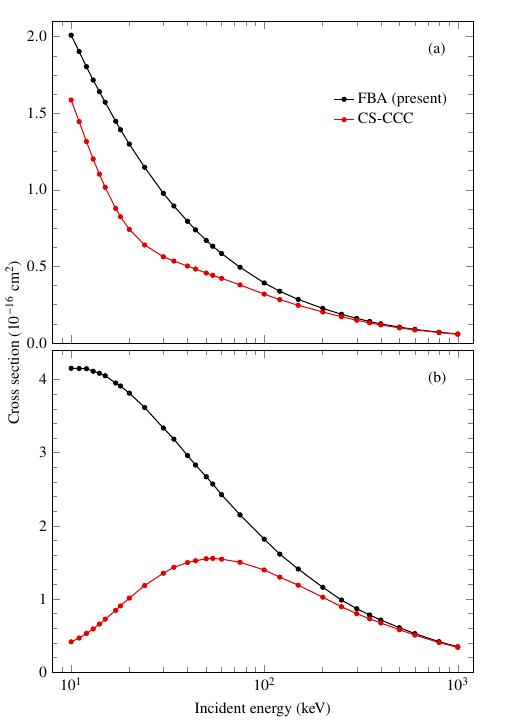}
    \caption{
        Cross sections for proton-impact excitation of the \ce{$2s 2p^2$ ^2D} 
        (a) and \ce{$2s 2p^2$ ^2P} (b) states of \ce{B}. The present 
        calculations using the CS-CCC and FBA methods are shown in both panels.
    }
    \label{fig4}
\end{figure}

\begin{figure}[t]
    \centering
    \includegraphics[width=\columnwidth]{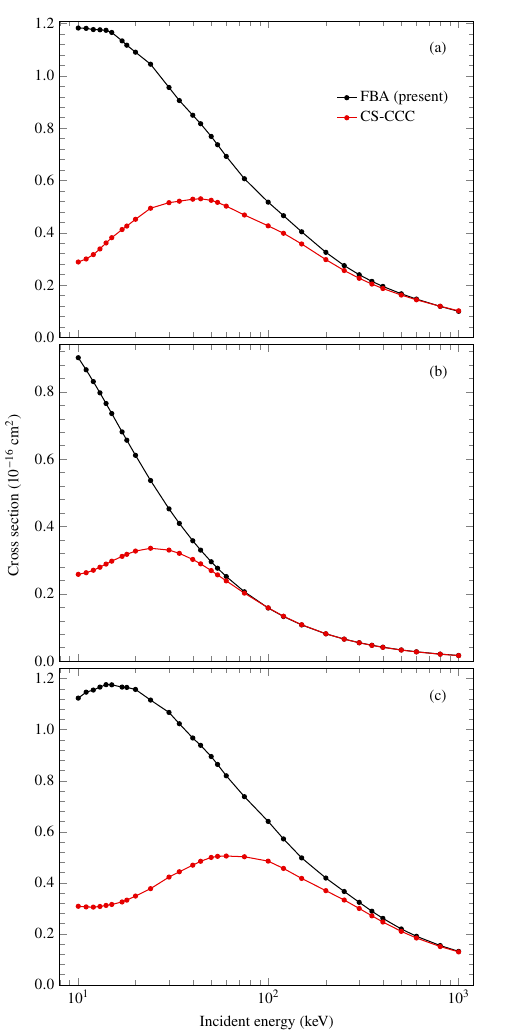}
    \caption{
        Cross sections for proton-induced excitation of the \ce{$2s^2 3s$ ^2S} 
        (a), \ce{$2s^2 3p$ ^2P^{\rm o}} (b), and \ce{$2s^2 3d$ ^2D} (c) 
        states of \ce{B}. The 
        present calculations using the CS-CCC and FBA methods are shown in 
        all panels.
    }
    \label{fig5}
\end{figure}

In Fig. \ref{fig4} we present cross sections for proton-impact excitation of 
the \ce{$2s 2p^2$ ^2D} and \ce{$2s 2p^2$ ^2P} states of \ce{B}. 
The first of these two states is the second excited state of \ce{B} and the 
other is an autoionising state which lies above the ionisation limit. These are 
representative transitions corresponding to excitation of the inner-valence 
electrons of \ce{B}. The present
CS-CCC and FBA results are shown in both panels. As one can see, the cross 
section corresponding to excitation of the \ce{$2s 2p^2$ ^2D} state falls 
rather sharply between 10 keV and 30 keV, while falling further but more slowly 
at higher incident energies. At 10 keV, excitation of this single state comprises 
about 44\% of the total excitation cross section. Such a large 
contribution from exciting a \ce{$2s$} subshell electron to the total 
excitation cross section emphasises the importance of having an accurate 
description of the multi-electron structure of the target
including the configurations with a hole in the $2s$ subshell. If one were to
apply an effective one-electron model to describe the outermost electron of 
\ce{B}, then a significant contribution to the total excitation cross section 
would not be accounted for.

\begin{figure}[tb]
    \centering
    \includegraphics[width=\columnwidth]{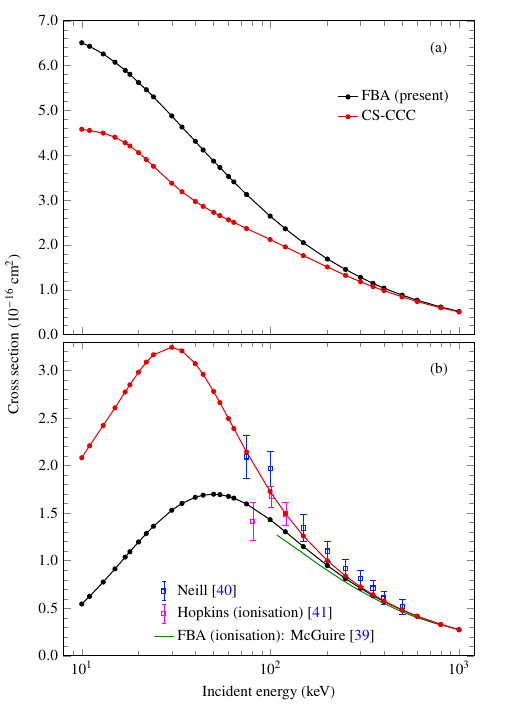}
    \caption{
        Cross sections for (a) total excitation and (b) electron loss 
        in \ce{$p$ + C^{+}} collisions. Present calculations 
        using the CS-CCC and FBA methods are shown in both panels. In 
        panel (b) the FBA cross section for total ionisation by 
        \citet{McGuire_1984_29} is shown. Additionally in the bottom panel, 
        the experimental measurements of the electron-loss cross section
        by \citet{Neill_1983_16} and the total ionisation cross section by
        \citet{Hopkins_1987_20} are also shown.
    }
    \label{fig6}
\end{figure}

Figure \ref{fig5} presents the cross sections for proton-impact excitation 
of the \ce{$2s^2 3s$ ^2S}, \ce{$2s^2 3p$ ^2P^{\rm o}}, and \ce{$2s^2 3d$ ^2D} 
states of \ce{B}. These states are examples of 
excitations of the outer-valence electrons of \ce{B}. Interestingly, at 10 keV
the cross sections for outer-electron excitation are about an order of 
magnitude smaller than the cross section for \ce{$2s 2p^2$ ^2D} excitation 
presented in Fig. \ref{fig4}. However, between 30 and 100 keV, these 
outer-valence excitation cross sections begin to peak and become comparable to 
the inner-valence electron excitation cross section. It is this behaviour of 
the outer-valence electron excitation cross section peaking at higher incident 
energies than the inner-valence electron excitation cross section that is 
responsible for the plateau seen in the total excitation cross section for 
\ce{$p$ + B} collisions in the middle panel of Fig. \ref{fig3} at incident 
energies between 20 and 40 keV. As is the case for the total cross 
sections presented in Figs. \ref{fig3} (and also in \ref{fig6} to be discussed 
in the next section), the FBA 
results converge to the corresponding CS-CCC ones at sufficiently high incident 
energies in both Figs. \ref{fig4} and \ref{fig5}. The fact that this 
occurs at the state-resolved level is a good test of the consistency of 
the two methods. 

\begin{figure}[t]
    \centering
    \includegraphics[width=\columnwidth]{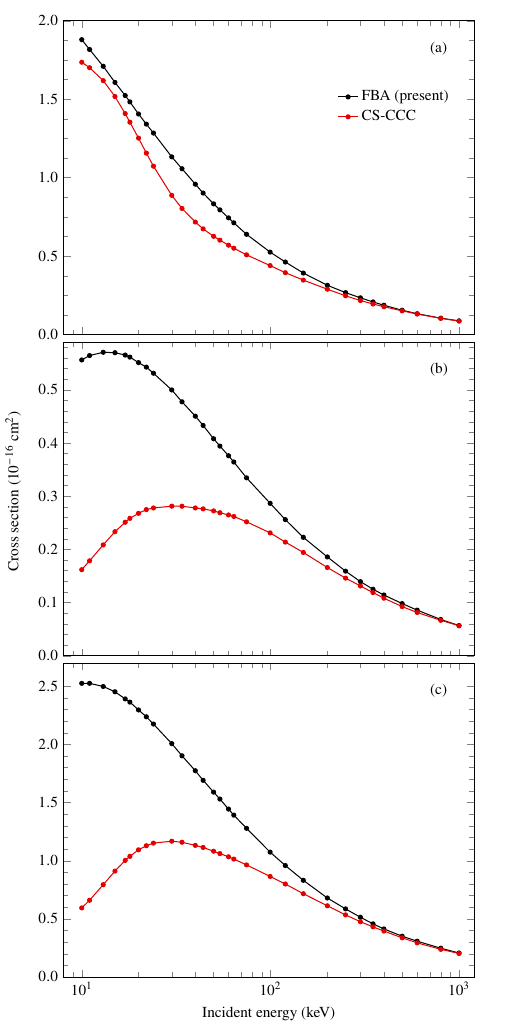}
    \caption{
        Cross sections for proton-induced excitation of the \ce{$2s 2p^2$ ^2D} 
        (a), \ce{$2s 2p^2$ ^2S} (b), and \ce{$2s 2p^2$ ^2P} (c) states of 
        \ce{C^{+}}. The 
        present calculations using the CS-CCC and FBA methods are shown in 
        all panels.
    }
    \label{fig7}
\end{figure}

\subsection{\ce{$p$ + C^{+}} collisions}

For \ce{$p$ + C^{+}} collisions, there are experimental 
measurements of the total electron capture and ionisation cross sections by 
\citet{Hopkins_1987_20} as well as separate measurements of the electron-loss
cross section by \citet{Neill_1983_16}. This makes studying collisions with 
\ce{C^+} particularly interesting as it allows the first direct comparison of 
our multi-electron CS-CCC results with experimental data. 
Additionally, \ce{C^+} is a boron-like ion, which means any conclusions drawn 
from the comparison of our results with experimental data for \ce{C^+} should 
indirectly support the validity of our method being applied to \ce{B} as well, 
for which there are no experimental data available for comparison. With these 
motivations in mind, we present the cross sections for total 
excitation and electron loss in \ce{$p$ + C^{+}} collisions in panels (a) and 
(b) of Fig. \ref{fig6}, respectively. In panel (b) we also 
show the FBA cross section for total ionisation by \citet{McGuire_1984_29}, as 
well as the experimental measurements of the electron-loss cross section by 
\citet{Neill_1983_16} and the total ionisation cross section by 
\citet{Hopkins_1987_20}. Note that for this system we do not present results 
for elastic scattering. This is because for collisions with charged targets, 
the elastic-scattering cross section is infinite due to the long-range Coulomb 
interaction between the projectile and the target \cite{Taylor_1972_00}. 

Comparison of the present CS-CCC calculations for the electron-loss cross 
section with the experimental measurements of \citet{Neill_1983_16} in panel 
(b) of Fig. \ref{fig6} shows excellent agreement over the entire overlapping 
incident-energy range between 80 and 500 keV. This is a very encouraging 
result, as it provides a strong validation of the present CS-CCC method for 
treating collisions with multi-electron targets. Assuming that the contribution 
of electron capture is already small at about 100 keV and becomes practically 
negligible above 200 keV, we compare our calculated electron-loss cross section 
with the total-ionisation cross-section measurements reported by 
\citet{Hopkins_1987_20}. We find very good agreement with the measurements
taken at 100 and 120 keV. However, there is a noticeable discrepancy with the 
measurement at 75 keV.

As was the case for \ce{$p$ + B} collisions, the present CS-CCC and FBA 
results for electron loss agree well with the FBA ionisation cross section of 
McGuire \cite{McGuire_1984_29} at sufficiently high incident energies. However, 
at lower energies the FBA results by McGuire also start deviating from the 
present FBA one, the former do not extend to energies below 100 keV. The reason 
for this deviation could be the same as mentioned above regarding the FBA by
\citet{Peach_1968_01,Peach_1970_03} in the case of \ce{$p$ + B} collisions.

\begin{figure}[b]
    \centering
    \includegraphics[width=\columnwidth]{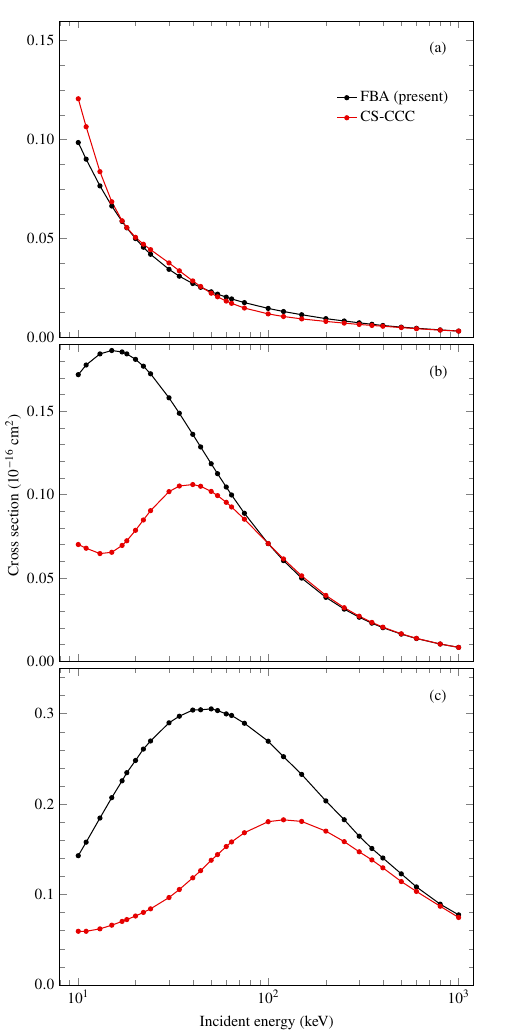}
    \caption{
        Cross sections for proton-induced excitation of the \ce{$2s^2 3s$ ^2S} 
        (a), \ce{$2s^2 3p$ ^2P^{\rm o}} (b), and \ce{$2s^2 3d$ ^2D} (c) 
        states of \ce{C^{+}}. The 
        present calculations using the CS-CCC and FBA methods are shown in 
        all panels.
    }
    \label{fig8}
\end{figure}

In Figs. \ref{fig7} and \ref{fig8} we present the cross sections for 
proton-impact excitation 
of some of the inner-valence and outer-valence excited states of \ce{C^+}, 
respectively. In Fig. \ref{fig7} we include 
excitation of the first three excited states of \ce{C^+}, which are all 
inner-valence excited states with a hole in the $2s$ subshell. In Fig.
\ref{fig8} we include excitation of the next three excited states of \ce{C^+}, 
following the ones in Fig. \ref{fig7}, which are all outer-valence excited 
states with a hole in the $2p$ subshell. Similar to what we observe for 
collisions with \ce{B}, Figs. \ref{fig7} and \ref{fig8} show that excitation of 
the $2s$ subshell electrons is dominant. However, unlike in the case of the 
\ce{B} target, this remains an equally important contribution to the total 
excitation cross section even at higher incident energies. This is perhaps not 
surprising, since the oscillator strength for the dipole-allowed transition 
\ce{$2s^2 2p$ ^2P^{\rm o}} $\to$ \ce{$2s 2p^2$ ^2P}, presented in 
Table \ref{tab:oscillator_strengths-C+}, is the largest among all 
dipole-allowed transitions from the ground state. Given that the cross section 
for excitation of a dipole-allowed transition becomes proportional to the 
corresponding oscillator strength at high incident energies 
\cite{Bransden_1992_00}, it is therefore straightforward to predict which 
excitation cross sections will dominate in the high-energy limit. This 
particular dipole-allowed transition is also the largest one for the \ce{B} 
target, as seen in Table \ref{tab:oscillator_strengths-B}. However, in the 
case of \ce{B}, the \ce{$2s 2p^2$ ^2P} state lies above the ionisation limit 
and therefore does not contribute to the total excitation cross section. In 
contrast, for \ce{C^+} it is a bound state and therefore does contribute to 
the total excitation cross section. This provides an example of how differences 
in the target structure and resulting spectrum can have a significant effect on 
the collision dynamics, despite both targets being boron-like. At the highest 
incident energy considered in this work, the electron-loss cross section is 
larger than the excitation cross section for \ce{B}, whereas the opposite 
holds for \ce{C^+}. This is due in part to the different role of the 
\ce{$2s 2p^2$ ^2P} state in the two targets.

\section{Conclusions and outlook} \label{sec:conclusions}
The single-centre CS-CCC method for ion collisions with multi-electron targets 
has been applied to proton collisions with \ce{B} and \ce{C^{+}} targets.
In addition to a comprehensive study of excitation and electron-loss 
processes, we have presented significant improvements to the construction of the 
target structure models by building configuration state functions using the 
method of coefficients of fractional parentage \cite{Cowan_1981_00}. This has 
allowed us to construct the target structure models in a much more systematic 
way than before, which is important for ensuring that the target structure 
remains accurate and robust as the number of electrons increases and that the 
relevant collision dynamics are properly captured.

Using a single-centre formalism to treat an inherently two-centre problem such 
as \ce{$p$ + B} and \ce{$p$ + C^{+}} collisions requires a very large 
pseudostate basis to reach convergence for the cross sections of 
interest in this work. This problem was further exacerbated by the fact that 
we considered collisions with multi-electron targets, which typically require a
larger pseudostate basis at the outset compared to performing coupled-channel 
calculations for one- \cite{Antonio_2024_110}
or two-electron targets \cite{Antonio_2025_112}. 
However, 
the significant performance improvements we were able to achieve with the GPU 
implementation of the CS-CCC code allowed us to include a sufficiently large 
pseudostate basis of 16000 pseudostates in the close-coupling calculations to 
ensure convergence of the cross sections of interest to within a few percent in 
general.

We presented cross sections for elastic scattering, total excitation and 
electron loss in \ce{$p$ + B} collisions, as well as cross sections for total excitation and 
electron loss in \ce{$p$ + C^{+}} collisions. This was done in the broad 
incident energy range of 10 keV to 1 MeV. 

For \ce{$p$ + C^{+}} collisions, we compared our calculations for 
the electron-loss cross section with experimental measurements by 
\citet{Neill_1983_16}. We found 
excellent agreement between our calculations and the corresponding experimental 
measurements over the entire overlapping incident energy range between 80 and 
500 keV. This is a very encouraging result as it provides a strong validation 
of the present CS-CCC method for treating collisions with multi-electron 
targets. We also compared our calculated electron-loss cross section with the 
total ionisation cross section measurements by \citet{Hopkins_1987_20} at 
incident energies where electron capture is expected to be negligible. We 
found very good agreement with the measurements taken at 100 and 120 keV, but 
a noticeable discrepancy with the measurement at 75 keV.
This could possibly be due to the growing contribution of the electron capture 
process to the total electron loss as the incident energy decreases.

For both systems, we compared our 
results for electron loss with previously calculated FBA cross sections for 
ionisation. In both cases, as expected, we found good agreement between our calculated 
electron-loss cross sections using both the CS-CCC and FBA methods with these
FBA ionisation cross sections at sufficiently high incident energies where 
electron capture is negligible. 

Throughout the set of results presented in this work, we observed significant 
differences between the CS-CCC and FBA results at lower incident energies, both 
in terms of the respective magnitudes of the cross sections and their 
incident energy dependence. This tells us that when interested in obtaining 
accurate cross sections at lower incident energies, it is paramount to account 
for the coupling between channels in the close-coupling equations, which is
not included in the FBA and other perturbative approaches. 

Overall, this work marks a significant step forward in the development of the 
CS-CCC method for treating ion-atom collisions involving arbitrary multi-electron 
targets. With the important methodological improvements made to the construction of 
the target structure models, together with the significant performance gains of 
the GPU implementation of the CS-CCC code, we are now in a very good position 
to consider applying the method to more complex processes, including various 
two-electron processes such as double ionisation and transfer excitation. A 
necessary next step toward this goal is to separate the contributions of 
electron capture and ionisation to the total electron-loss cross section so 
that state-resolved electron-capture cross sections can be obtained. This is 
important because such data are needed for applications in fusion-plasma 
impurity diagnostics and modelling. This can be achieved in two ways. The first 
would be to extend the present single-centre formalism to a two-centre 
formalism following a procedure similar to that used in the two-centre WP-CCC 
method \cite{Abdurakhmanov_2018_97}. This would allow us to explicitly include 
the bound and continuum states of the projectile in the close-coupling 
expansion and thereby obtain state-resolved electron-capture cross sections 
directly. The second would be to apply the projection method introduced by 
\citet{Abdurakhmanov_2020_53}, which would allow us to utilise the existing 
single-centre formalism and the highly optimised code already in place, while 
still separating the contributions of electron capture and ionisation in an 
\textit{ab initio} way.

\begin{acknowledgments}
    This work was supported by the Australian Research Council (Grant No.
    DP240101210). We acknowledge the resources provided by Pawsey 
    Supercomputing Centre. N. W. Antonio and A. S. Kadyrov would like to 
    acknowledge the support provided by the Pawsey Supercomputing Research 
    Centre through an Uptake project dedicated to the development of 
    the GPU implementation of the CS-CCC code. 
\end{acknowledgments}

%

\end{document}